\documentclass[aps,pra,twocolumn,nofootinbib,longbibliography,10pt]{revtex4}
\usepackage{comment}
\usepackage{hyperref}
\usepackage{graphicx} 
\usepackage{xcolor}
\usepackage{amsmath}
\usepackage{amsfonts}
\usepackage{amssymb}
\usepackage{xcolor}
\usepackage{graphicx}
\usepackage{dcolumn}
\usepackage{bm}
\usepackage{hyperref,etoolbox}
\usepackage{mathrsfs}
\usepackage{subfig}
\usepackage{natbib}
\usepackage{caption}
\usepackage[none]{hyphenat}
\newcommand{\ket}[1]{\left|#1\right>}
\newcommand{\bra}[1]{\left<#1\right|}
\newcommand{\braket}[2]{\left<#1\mid#2\right>}

\begin{document}
\title{Current, \textcolor{black}{quantum transport} and entropic force of bosonic system interacting with two thermal reservoirs}
\author{\textbf{Jayarshi Bhattacharya}}
\email{dibyajayarshi@gmail.com}
\author{\textbf{Gautam Gangopadhyay}}
\email{gautam@bose.res.in}
\author{\textbf{Sunandan Gangopadhyay}}
\email{sunandan.gangopadhyay@gmail.com}
\affiliation{\textit{S.N. Bose National Centre for Basic Sciences, JD Block, Sector-III, Salt Lake, Kolkata 700106, India}}
\begin{abstract}
	\noindent This paper investigates the dynamics of current and \textcolor{black}{quantum transport factor} in a bosonic system consisting of a central system interacting with two reservoirs at different temperatures. We derive a master equation describing the time evolution of the density matrix of the system, accounting for the interactions and energy transfer between the components. We quantify the current, representing the flow of bosons through the system and analyse its dependence on the system's parameters and temperatures of the thermal reservoirs. In the steady state regime, we derived an expression for the \textcolor{black}{quantum transport factor} of the energy transfer process. Our analysis show that quantum effects, such as the dependence on temperature can significantly impact \textcolor{black}{this factor. In particular, we observe that the transport factor of the quantum system is greater than the corresponding factor when the temperature goes to infinity, where the factor has an identical form with the Carnot efficiency of an ideal heat engine}. We then derived the Fokker-Planck equation to find out the Glauber-Sudarsan $P$-representation. {In the steady state of the equation, the probability distribution comes out to be in Gaussian form. We then calculated the entropic force for this probability distribution which gives the Hooke's law in the steady state, in agreement with the fact that our system is a harmonic oscillator.}
\end{abstract}

\maketitle
\section{Introduction}
\noindent Advancement in energy transport factor has become an important prime trait in both classical and quantum thermodynamics that has drawn great attention from the scientific community by making dramatic strides in technology \cite{cakmak2020construction,PhysRevE.76.031105,unknown2023comparative}. The study in quantum thermodynamics is now a very important field to study in general \cite{abe2011similarity,gemmer2009quantum,alicki2018introduction}. The discovery of fundamental limits in the process of energy conversion has paved a way for great insight and many out-of-the-box possibilities, especially in this intricate domain of quantum systems. This work makes an in-depth exploration of current and \textcolor{black}{quantum transport factor} in a bosonic system, shedding light on the intricate dance of bosons inside one central system and their dynamic interaction with reservoirs operating at different temperatures. This quest is highly motivated by the potential to exploit quantum mechanical effects to optimize the transfer and utilization of energy, Thus, opening new frontiers in quantum optics \cite{Glauber1963, sudarshan1963equivalence, Carmichael, Louisell, Zubairy}.\\
In quantum optics, the open quantum systems have been studied rigorously in \cite{weinbub2023computational, shaker2023advancements}. A system in attached thermal bath has always been a point of interest \cite{Mukherjee:2024egj} for the physicists. In this paper, we will concentrate on the ensemble of interacting bosons involving two reservoirs, {an emitter attached to a reservoir at temperature $T_e$, and a collector attached to another reservoir at temperature $T_c$}. One of the features that makes this model rather interesting is the involvement of bosons in both the central system and the reservoirs. The solution to this complicated problem begins with the construction of the system's Hamiltonian in such an extremely elegant way that a central system couples to several reservoirs \cite{Cahill1969}. Subsequently, we meticulously derive the master equation that governs the density matrix of the central system, intricately accounting for the dynamic interactions and energy transfer mechanisms operating between the constituent elements. {However, non-Markovian systems have also been studied in \cite{PhysRevB.92.235440, PhysRevB.95.064308, PhysRevD.45.2843}. In these articles different types of open quantum systems have been investigated to derive exact master equations. For example, in \cite{PhysRevB.92.235440} a single oscillator coupled with an electron bath was studied, and the heat flow for the system was obtained. In this article we concentrate only on the Markovian system to find some useful results.}\\
{Our focus in this article is unraveling the quantum mechanical behavior of current and \textcolor{black}{quantum transport factor} in a bosonic system.} We consider the detailed analysis of current quantification, pointing out its flow through the system, where it becomes really sensitive to many parameters of the system and temperatures. We discuss the role that quantum mechanical effects play in shaping current dynamics, investigating how the quantum nature of bosons influences the flow within the system. Diving deep into the study, there's a sharp focus on how well energy moves around, especially when things balance out and hit a steady rhythm. Think of it like tracking the path of water flowing through a series of pipes; \textcolor{black}{quantum transport factor} shows how smoothly it flows without leaking or getting blocked up.\\
In the subsequent sections, we unfold the mathematical formalism that acts as the bedrock for calculating both the current and \textcolor{black}{quantum transport factor} within the bosonic system. Then we delve into deriving the coherent {state} representation that plays a crucial role for this system. {We derive the Fokker-Planck equation using the  Glauber-Sudarshan $P$-representation. The Fokker-Planck equation is then solved to get the solution of $P$-representation. This probability distribution happens to be in Gaussian form in the steady state. Then from the probability distribution, we calculate the entropic force \cite{verlinde2011origin, roos2014entropic, bhattacharya2023entropic} to get more insight about the steady state condition in non-equilibrium thermodynamic systems.}\\
Furthermore, our journey delves into the far reaching implications of our findings, emphasizing their significance in the broader canvas of quantum thermodynamics. These insights contribute to promise for practical applications, potentially influencing the development of quantum technologies with enhanced energy utilization capabilities.\\
This article is organised as follows. In section \ref{sec1}, we discussed about defining the current which leads us to define the quantum transport factor in section \ref{sec2}.  In section\ref{sec3}, we discussed the Fokker-Planck equation for our system and then calculated the entropic force of the system. Finally, we concluded in section \ref{sec4}. 

\section{Current for Bosons}\label{sec1}
\noindent In this study, we explore the intricate dynamics of a system of bosons interacting with two distinct reservoirs, an emitter and a collector, each possessing its own temperature, denoted as $T_e$ and $T_c$, respectively. It is noteworthy that these reservoirs are also bosonic in nature. We can represent this system through the following Hamiltonian.
\begin{subequations}\label{hamiltonian}
	\begin{align}
		\hat{H_s}&=\hbar\omega_s \hat{a}^\dagger \hat{a}\\
		\hat{H_e}&=\hbar\sum\limits_{j}\omega^e_j \hat{b_j}^\dagger \hat{b_j}\label{1b}\\
		\hat{H_c}&=\hbar\sum\limits_{j}\omega^c_j \hat{d_j}^\dagger \hat{d_j}\label{1c}\\
		\hat{H_i}&=\hbar\sum\limits_{j}\left( \Gamma_{ej} \hat{a}^\dagger \hat{b_j} +\Gamma_{ej}^* \hat{a} \hat{b}^\dagger_j +\Gamma_{cj} \hat{a}^\dagger \hat{d_j} +\Gamma_{cj}^* \hat{a} \hat{d_j}^\dagger \right)
	\end{align}
\end{subequations}
where $\hat{a}$, $\hat{b_j}$, and $\hat{d_j}$ are operators obeying the bosonic commutation relation $\left[\hat{b_j},\hat{b_k}^\dagger\right]=\delta_{jk}=\left[\hat{d_j},\hat{d_k}^\dagger\right]$ and $\left[\hat{a},\hat{a}^\dagger\right]=1$. \textcolor{black}{The above Hamiltonian of a single harmonic oscillator coupled to two reservoirs at two different temperatures is physically relevant in the context of quantum transport of particles. In particular the study of transport properties of systems with a discrete energy level have been very important \cite{postma2001carbon,datta1997electronic,liang2002kondo}, and it is here that the model considered in this work finds its relevance. The quantum current that we shall discuss later can be realised by an arrangement of source-system-sink, with the source and sink being bosonic reservoirs whose Hamiltonians are given in eq.(s) \eqref{1b} and \eqref{1c}. For the sake of completeness, we would like to mention that the problem of a single harmonic oscillator coupled to a reservoir is also relevant in quantum optics scenarios. The reservoir oscillators may represent the Fourier modes of the radiation field into which an excited atom can decay through the process of spontaneous emission \cite{Carmichael}.}\\
To proceed with our investigation, we adopt the density matrix formalism in the interaction picture. The time evolution of the density matrix $\hat{\rho_s}$ in interaction picture can be described by the following differential equation \cite{Carmichael,Zubairy,Louisell}

\begin{align}\label{diffeq}
	\frac{d\hat{\rho_s}}{dt}&=-\int_0^{t}dt' ~~\text{Tr}_{e,c}\left[\hat{H_i}(t),\left[\hat{H_i}(t'),\hat{\rho}(t')\right]\right]~.
\end{align}
where, $\hat{\rho}(t')=\hat{\rho_s}(t')\otimes\hat{\rho_e}(0)\otimes\hat{\rho_c}(0)$.\\
This equation accounts for the intricate interplay between the system and its environment, where $\hat{H_i}(t)$ is the interaction Hamiltonian. It is instrumental in unraveling the dynamics of our bosonic system.\\
Following the prescription in \cite{Carmichael}, one can write the master equation of the system in interaction picture as
\begin{align}
	\frac{d\hat{\rho}_s}{dt}&=-i\Delta\left[\hat{a}^\dagger \hat{a},\hat{\rho}_s\right]- \frac{\gamma_e+\gamma_c}{2}\left(\hat{a}^\dagger\hat{a}\hat{\rho}_s+\hat{\rho}_s\hat{a}^\dagger \hat{a}-2\hat{a}\hat{\rho}_s\hat{a}^\dagger\right)\nonumber\\
	&\hspace{6mm}+ \left(\gamma_e\bar{n}_e+\gamma_c\bar{n}_c\right)\left(\hat{a}\hat{\rho}_s\hat{a}^\dagger+\hat{a}^\dagger\hat{\rho}_s\hat{a}-\hat{\rho}_s\hat{a}\hat{a}^\dagger-\hat{a}^\dagger\hat{a}\hat{\rho}_s\right)~.
\end{align}
In this equation, $\Delta$ represents the frequency shift while $\gamma_e$ and $\gamma_c$ represent the damping coefficients, which are defined as
\begin{subequations}\label{gamma}
	\begin{align}
		\Delta&=P\int_0^\infty d\omega\frac{g(\omega)\left(\left|\Gamma_e(\omega)\right|^2+\left|\Gamma_c(\omega)\right|^2\right)}{\omega_s-\omega}\\
		\gamma_e&=2\pi g(\omega_s)\left|\Gamma_{e}(\omega_s)\right|^2\\
		\gamma_c&=2\pi g(\omega_s)\left|\Gamma_{c}(\omega_s)\right|^2
	\end{align}
\end{subequations}
{where $g(\omega)$ is the density of states arising in going to the continuum from the discrete case, and $P$ is the principal value of the integral. $\Gamma_j$'s are Thus, replaced with $\Gamma(\omega)$'s in the continuum case.}\\
Then the time evolution of the density matrix $\hat{\rho}$ in the Schroedinger picture is given by
\begin{align}
	\frac{d\hat{\rho}}{dt}&=-i\omega_s\left[\hat{H}_s,\hat{\rho}\right]-e^{-i/\hbar\hat{H}_s t}{\dot{\hat{\rho}}_s} e^{i/\hbar\hat{H}_s t}\nonumber\\
	&=-i\omega\left[\hat{a}^\dagger \hat{a},\hat{\rho}\right]- \frac{\gamma_e+\gamma_c}{2}\left(\hat{a}^\dagger\hat{a}\hat{\rho}+\hat{\rho}\hat{a}^\dagger \hat{a}-2\hat{a}\hat{\rho}\hat{a}^\dagger\right)\nonumber\\
	&\hspace{6mm}+ \left(\gamma_e\bar{n}_e+\gamma_c\bar{n}_c\right)\left(\hat{a}\hat{\rho}\hat{a}^\dagger+\hat{a}^\dagger\hat{\rho}\hat{a}-\hat{\rho}\hat{a}\hat{a}^\dagger-\hat{a}^\dagger\hat{a}\hat{\rho}\right)\label{masteq}
\end{align}
where $\omega=\omega_s+\Delta$.\\
These coefficients play a pivotal role in quantifying the dissipative processes within the system. Furthermore, $\bar{n}_e$ and $\bar{n}_c$ denote the number densities, defined as
\begin{subequations}\label{n}
	\begin{align}
		\bar{n}_e&=\frac{1}{e^{\frac{\hbar\omega_s}{k_B T_e}}-1}\\
		\bar{n}_c&=\frac{1}{e^{\frac{\hbar\omega_s}{k_B T_c}}-1}~.
	\end{align}
\end{subequations}
These number densities reflect the equilibrium distributions of bosons within the emitter and collector reservoirs.\\
In our pursuit of understanding the dynamics of this bosonic system, it is important to investigate the average number of bosons within the system. This quantity, denoted as $\left<\hat{n}(t)\right>$, can be derived by examining the time evolution of the density matrix of the system. This gives,
\begin{align}\label{number}
	\left<\dot{\hat{n}}\right>(t) &=\left<\hat{a}^\dagger\hat{a}\dot{\hat{\rho}}(t)\right>\nonumber\\
	&=-\left(\gamma_e+\gamma_c\right)\left(\left<\hat{n}\right>(t)-\frac{\gamma_e\bar{n}_e+\gamma_c\bar{n}_c}{\gamma_e+\gamma_c}\right)
\end{align}
Solving this differential equation gives
\begin{align}
	\left<\hat{n}\right>(t)&= \frac{\gamma_e\bar{n}_e+\gamma_c\bar{n}_c}{\gamma_e+\gamma_c} +\nonumber\\
	&~~\left(\left<\hat{n}(0)\right>-\frac{\gamma_e\bar{n}_e+\gamma_c\bar{n}_c}{\gamma_e+\gamma_c}\right)e^{-(\gamma_e+\gamma_c)t}\label{extra1}~.
\end{align}
This elegant expression reveals how the average number of bosons evolves over time, taking into account the damping effects caused by $\gamma_e$ and $\gamma_c$. It also incorporates the initial state of the system, as represented by $\left<\hat{n}(0)\right>$.\\
Moving forward, let us consider the steady state number density, denoted as $\bar{n}_s$. This quantity provides valuable insights into the long-term behavior of the system. The steady state number can be obtained by setting $\left<\dot{\hat{n}}\right>(t)=0$ in eq. \eqref{number}. This gives
\begin{align}
	\bar{n}_s=\frac{\gamma_e\bar{n}_e+\gamma_c\bar{n}_c}{\gamma_e+\gamma_c}\label{steady_number}~.
\end{align}
It is interesting to note that the above result can be recast in the form
\begin{align}
	\gamma_e(\bar{n}_e-n_s)=\gamma_c(n_s-\bar{n}_c)\label{balance}~.
\end{align}
Eq. \eqref{steady_number} gives an expression for the steady state number density, showcasing how the system reaches equilibrium. Eq. \eqref{balance} further underscores the balance achieved in the system between the emitter and collector reservoirs, with $\gamma_e$ and $\gamma_c$ playing pivotal roles. Eq. \eqref{balance} can be interpreted as the flux balance in the steady state condition.\\
Again, in the steady state, that is, when $\dot{\hat{\rho}}=0$, from eq.\eqref{masteq} and eq.\eqref{steady_number}, we can write,
\begin{align}
	&n\rho_{nn}-(n+1)\rho_{n+1,n+1}\nonumber\\
	&\hspace{8mm}=\bar{n}_s\left\{(n+1)\rho_{n+1,n+1}+n\rho_{n-1,n-1}-(2n+1)\rho_{nn}\right\}\nonumber\\
	&\text{or,}\hspace{2mm}(\bar{n}_s+1)(n+1)\rho_{n+1,n+1}+n\bar{n}_s\rho_{n-1,n-1}\nonumber\\
	&\hspace{3cm}=\{n+\bar{n}_s(2n+1)\}\rho_{nn}
\end{align}
where $\rho_{nn}=\bra{n}\hat{\rho}\ket{n}$. This equation clearly gives the diagonal density matrix to be
\begin{align}
	\rho_{nn}&=\left(\frac{\bar{n}_s}{\bar{n}_s+1}\right)^n\rho_{00}\nonumber\\
	&=\frac{1}{\bar{n}_s+1}\left(\frac{\bar{n}_s}{\bar{n}_s+1}\right)^n
\end{align}
where we used $Tr(\hat{\rho})=1$ in the above result. This result resembles the bosonic distribution with average number density given in eq.\eqref{steady_number}.\\
Now we move on to the concept of current within this system. The left-hand side of eq. \eqref{balance}, that is, $\gamma_e(\bar{n}_e-\bar{n}_s)$ can be expressed as $\frac{\gamma_e\gamma_c}{\gamma_e+\gamma_c}(\bar{n}_e-\bar{n}_c)$. This term reveals a fundamental relationship between the imbalance in the emitter and collector reservoirs and the quantum current. In particular, the factor $\frac{\gamma_e\gamma_c}{\gamma_e+\gamma_c}$ is a well-established result in the literature, as found in references \cite{karmakar2016fermionic,sun1999quantum}. The above relation can be interpreted as the steady state current. {This leads us to define the average particle current for this system as follows}
\begin{align}
	\textcolor{black}{I(t)}&=\frac{1}{2}\left[\gamma_e\left({\bar{n}_e}-\left<\hat{n}\right>(t)\right)+\gamma_c\left(\left<\hat{n}\right>(t)-{\bar{n}_c}\right)\right]\label{current}~.
\end{align}
{Substituting $\left<\hat{n}\right>(t)$ from eq. \eqref{extra1} in the above equation leads to
\begin{align}
	\textcolor{black}{I(t)}&=\textcolor{black}{I_s}+\left(\textcolor{black}{I_0}-\textcolor{black}{I_s}\right) e^{-\left(\gamma_e+\gamma_c\right)t}\label{current_extra1}
\end{align}
where $\textcolor{black}{I_0}$ and $\textcolor{black}{I_s}$ are given by
\begin{align}
	\textcolor{black}{I_0}&=\frac{1}{2}\left[\gamma_e\left({\bar{n}_e}-\left<\hat{n}\right>(0)\right)+\gamma_c\left(\left<\hat{n}\right>(0)-{\bar{n}_c}\right)\right]~;\label{i0}\\
	\textcolor{black}{I_s}&= \frac{\gamma_e\gamma_c}{\gamma_e+\gamma_c}(\bar{n}_e-\bar{n}_c)\label{is}~.
\end{align}
This matches exactly with the balanced flux derived earlier.}\\
The current given by the eq. \eqref{current}, embodies the interaction between the emitter and the collector, reflecting how bosons flow within the system. The time derivative of the current, $\textcolor{black}{\frac{d I}{dt}}$, is given by,
\begin{align}
	\textcolor{black}{\frac{d I}{dt}}&=\frac{\left(\gamma_e-\gamma_c\right)\left(\gamma_e+\gamma_c\right)}{2}\left(\left<\hat{n}\right>(t)-\frac{\gamma_e\bar{n}_e+\gamma_c\bar{n}_c}{\gamma_e+\gamma_c}\right)\nonumber\\
	&=-\left(\gamma_e+\gamma_c\right)\left(\textcolor{black}{I(t)}-\frac{\gamma_e\gamma_c}{\gamma_e+\gamma_c}(\bar{n}_e-\bar{n}_c)\right)~.
\end{align}
This equation elucidates how the current changes over time and underscores the intricate interplay between various parameters that govern the system's behavior. Setting $\textcolor{black}{\frac{d I}{dt}}=0$ gives the current at the steady state to be $\textcolor{black}{I_s}$ as given in eq. \eqref{is}.\\
This expression quantifies the equilibrium flow of bosons within the system, reflecting the balance between the emitter and collector reservoirs.
\section{quantum transport factor}\label{sec2}
\begin{figure}[ht]
	\includegraphics[width=0.9\linewidth]{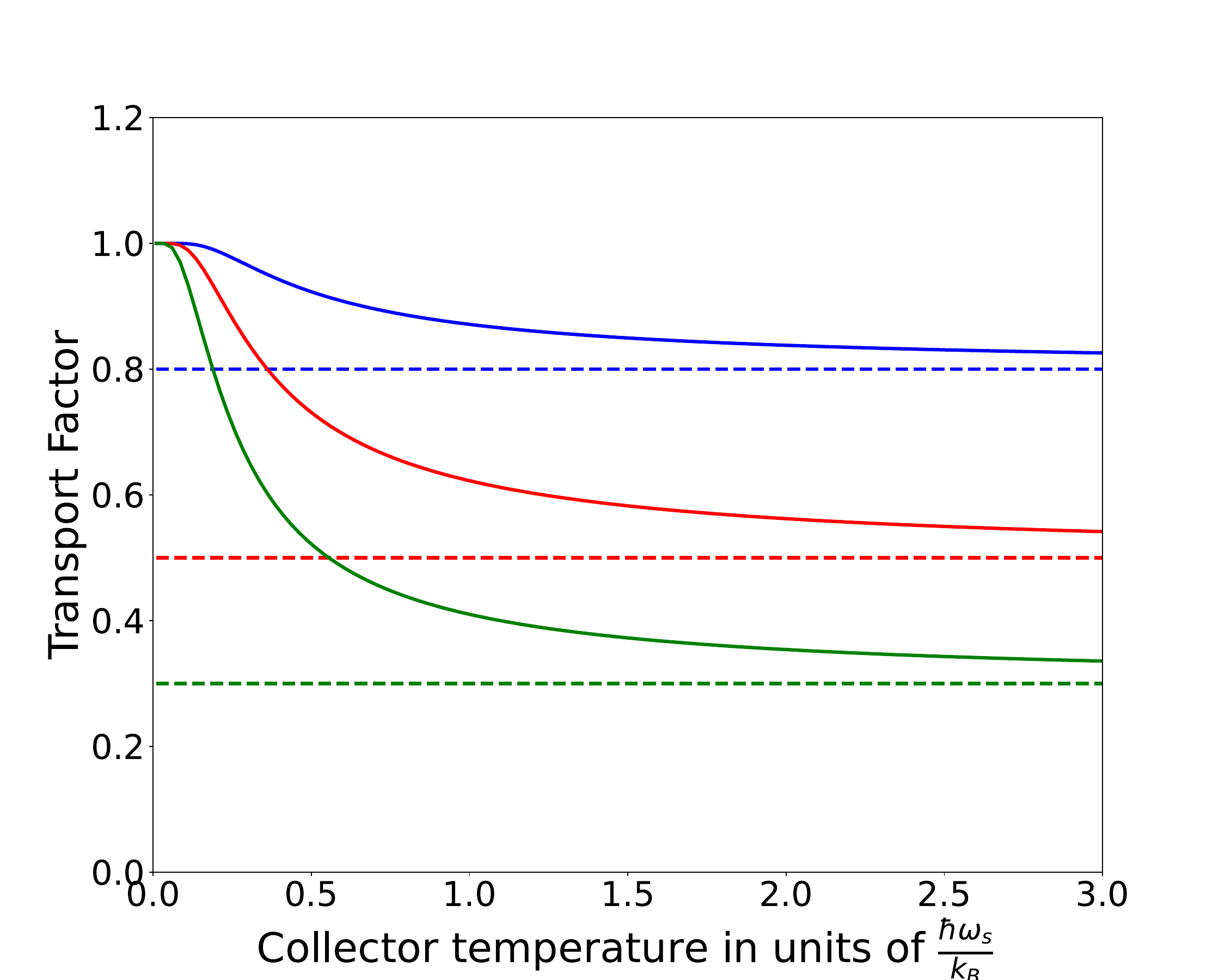}
	\caption{Steady state \textcolor{black}{quantum transport factor} is plotted against collector temperature measured in units of $\frac{\hbar\omega_s}{k_B}$ different temperatures.}
	\label{fig:fig1}
\end{figure}

\begin{figure}[ht]
	\includegraphics[width=0.9\linewidth]{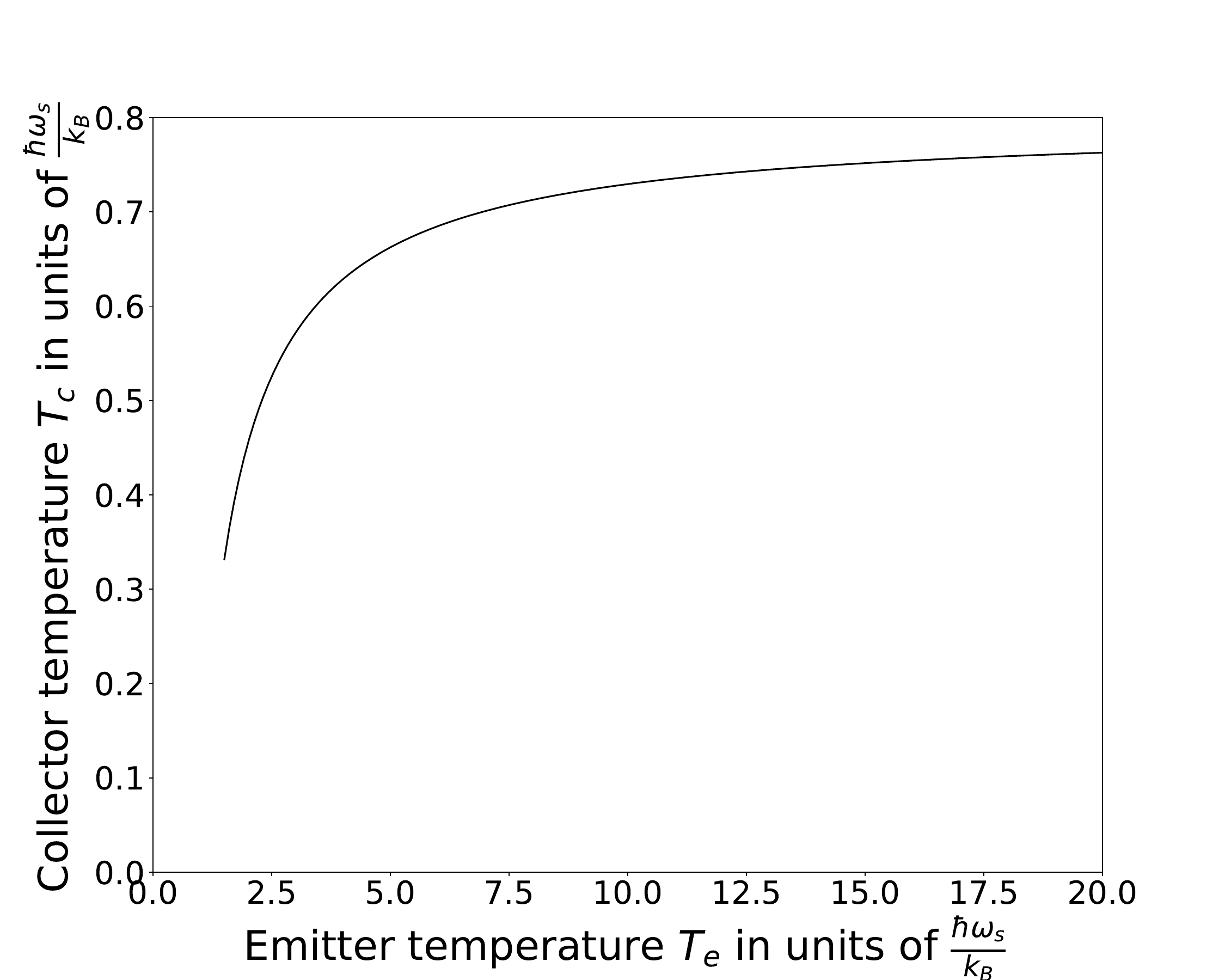}
	\caption{Plot of collector temperature $T_c$ vs emitter temperature $T_e$ when the \textcolor{black}{quantum transport factor} is half.}
	\label{fig:extra1}
\end{figure}

We now explore the concept of steady state energy loss per unit time, represented as $E_s$. This energy loss is a crucial parameter that characterizes the system's behavior. It is given by
\begin{align}
	E_s&=\hbar\omega_s \textcolor{black}{I_s}\nonumber\\
	&=\hbar\omega_s \frac{\gamma_e\gamma_c}{\gamma_e+\gamma_c}(\bar{n}_e-\bar{n}_c)~.
\end{align}
$E_s$ represents the energy lost from the system to the environment in equilibrium. It is given by the product of the energy of the harmonic oscillator representing the system ($\hbar\omega_s$) and the steady state current ($I_s$).\\
Moving on, let us consider the amount of energy the system recieves from the emitter at unit time, denoted as $Q$. This quantity can be expressed as
\begin{equation}
	Q=f(\gamma_e,\gamma_c)\hbar\omega_s \bar{n}_e
\end{equation}
{Here, \textcolor{black}{the factor $f(\gamma_e,\gamma_c)$ is needed from purely dimensional consideration and} has the dimension of $time^{-1}$}. $Q$ represents the energy loss per unit time from the emitter due to the physical processes within the system. It is proportional to the energy of the bosons in the emitter reservoir.\\
The \textcolor{black}{quantum transport factor} parameter $\eta_s$ provides valuable insights into how effectively the system performs \textcolor{black}{in transporting particles from source to sink}. It is defined as
\begin{align}
	\eta_s&= \frac{E_s}{Q}\nonumber\\
	&= \frac{1}{f(\gamma_e,\gamma_c)}\frac{\gamma_e\gamma_c}{\gamma_e+\gamma_c} \left(1-\frac{\bar{n}_c}{\bar{n}_e}\right)\label{eff1}~.
\end{align}
We now look at the high temperature limit of the above expression. From eq.(s) (\eqref{n}, \eqref{eff1}), we can express the \textcolor{black}{quantum transport factor} at high temperatures as
\begin{align}
	\eta_s &= \frac{1}{f(\gamma_e,\gamma_c)}\frac{\gamma_e\gamma_c}{\gamma_e+\gamma_c} \left(1-\frac{e^{\frac{\hbar\omega_s}{k_B T_e}}-1}{e^{\frac{\hbar\omega_s}{k_B T_c}}-1}\right)\nonumber\\
	&\approx \frac{1}{f(\gamma_e,\gamma_c)}\frac{\gamma_e\gamma_c}{\gamma_e+\gamma_c} \left(1-\frac{T_c}{T_e}\right)\label{eff2}\\
	&= \frac{1}{f(\gamma_e,\gamma_c)}\frac{\gamma_e\gamma_c}{\gamma_e+\gamma_c} \eta_{c}
\end{align}
where \textcolor{black}{the factor $\eta_{c}$ has a form exactly identical to the well-known Carnot efficiency} given by
\begin{align}
	\eta_c=1-\frac{T_c}{T_e}\label{carnot_effi}~.
\end{align}
Our next step is to fix the form of $f(\gamma_e,\gamma_c)$. This can be done by demanding that the above result for $\eta_s$ should be equal to $\eta_c$ in the high temperature limit. This immediately fixes $f(\gamma_e,\gamma_c)$ to be equal to $\frac{\gamma_e\gamma_c}{\gamma_e+\gamma_c}$. Consequently, the steady state \textcolor{black}{quantum transport factor} can be written as
\begin{align}
	\eta_s&=1-\frac{\bar{n}_c}{\bar{n}_e}\label{eff_steady}~.
\end{align}
{This clearly shows that the \textcolor{black}{quantum transport factor} is bounded from above by unity.} The time dependent general result of \textcolor{black}{quantum transport factor} can be written as
\begin{align}
	\eta(t)&=\frac{\textcolor{black}{I(t)}}{\frac{\gamma_e\gamma_c}{\gamma_e+\gamma_c}\bar{n}_e}\nonumber\\
	&=\frac{\gamma_e+\gamma_c}{\gamma_e \gamma_c}\frac{\textcolor{black}{I_s}}{\bar{n}_e}+\frac{\gamma_e+\gamma_c}{\gamma_e \gamma_c}\frac{\left(\textcolor{black}{I_0}-\textcolor{black}{I_s}\right)}{\bar{n}_e}e^{-\left(\gamma_e+\gamma_c\right)t}\nonumber\\
	&=1-\frac{\bar{n}_c}{\bar{n}_e}+\left[\frac{\gamma_e+\gamma_c}{\gamma_e\gamma_c}\frac{\textcolor{black}{I_0}}{\bar{n}_e} -\left(1-\frac{\bar{n}_c}{\bar{n}_e}\right)\right]e^{-(\gamma_e+\gamma_c)t}\label{efft}~.
\end{align}
At the steady state, that is, $t\to\infty$, the \textcolor{black}{quantum transport factor} $\eta_s$ as a function of temperature can be written using the result from eq.(s) \eqref{n} as follows
\begin{align}
	\eta_s &= 1-\frac{e^{\frac{\hbar\omega_s}{k_B T_e}}-1}{e^{\frac{\hbar\omega_s}{k_B T_c}}-1}\nonumber\\
	&=1-\left[\left(\frac{\hbar\omega_s}{k_B T_e}\right)+\frac{1}{2}\left(\frac{\hbar\omega_s}{k_B T_e}\right)^2+\cdots \right]\nonumber\\ &\hspace{10mm}\times\frac{k_B T_c}{\hbar\omega_s}\left[1+\frac{1}{2}\left(\frac{\hbar\omega_s}{k_B T_c}\right)+\cdots \right]^{-1}\nonumber\\
	&\approx 1-\frac{T_c}{T_e}+\frac{\hbar\omega_s}{2k_B T_e}\left(1-\frac{T_c}{T_e}\right)+\cdots\nonumber\\
	&={1-\frac{T_c}{T_e}+\frac{\hbar\omega_s}{2k_B T_c}\left(1-\frac{T_c}{T_e}\right)\frac{T_c}{T_e}+\cdots}\nonumber\\
	&=\eta_c+\frac{\hbar\omega_s}{2k_B T_c}\eta_c(1-\eta_c)+\cdots
\end{align}
{where we have used $\frac{\hbar\omega_s}{k_B T_e},\frac{\hbar\omega_s}{k_B T_c}\ll 1$ in writing down the second line of the above equality.}\\
{This result shows a deviation from the previously known results of quantum engines where the \textcolor{black}{quantum transport factor} is found to be equal to the classical results as shown in \cite{rezek2006irreversible,Mukherjee:2022prn,wang2009performance}. In general, it has been observed in \cite{moreira2020enhancing,li2022quantum} that the \textcolor{black}{quantum transport factor} in non-Markovian systems are greater than the Markovian case. In this analysis, which is based on Markovian approximation,} we found that the \textcolor{black}{\textcolor{black}{quantum transport factor} is indeed higher than the $\eta_c$ and agrees with $\eta_c$ when $T_c$ goes to infinity. As mentioned before, $\eta_c$ matches exactly with the Carnot efficiency.} This equation beautifully captures how the \textcolor{black}{quantum transport factor} of the system varies with temperature, demonstrating the impact of quantum corrections on \textcolor{black}{the factor}, which is evident from the temperature-dependent term in the summation. Remarkably, it can be seen from the above expression that the \textcolor{black}{transport factor} of the quantum system is greater than the $\eta_c$. In Fig. \ref{fig:fig1}, it is shown how \textcolor{black}{transport factor} increases for quantum systems. It is worth noting that in our analysis $T_c<T_e$.
\newline\noindent As the emitter temperature (in units of $h\omega_s/k_B$) increases, the collector temperature (in units of $\hbar\omega_s/k_B$) for which, the \textcolor{black}{quantum transport factor} becomes half of its original value increases and gradually gets saturated. The half \textcolor{black}{quantum transport factor} is then defined as $\frac{1}{2}(1-\eta_c)$ because the maximum value of $\eta_s$ is $1$. It is clearly shown in Fig. \ref{fig:extra1}.
\section{$P$-Representation and Fokker-Planck equation for bosonic system}\label{sec3}
\begin{figure*}[ht]
    \centering
    \includegraphics[width=\textwidth]{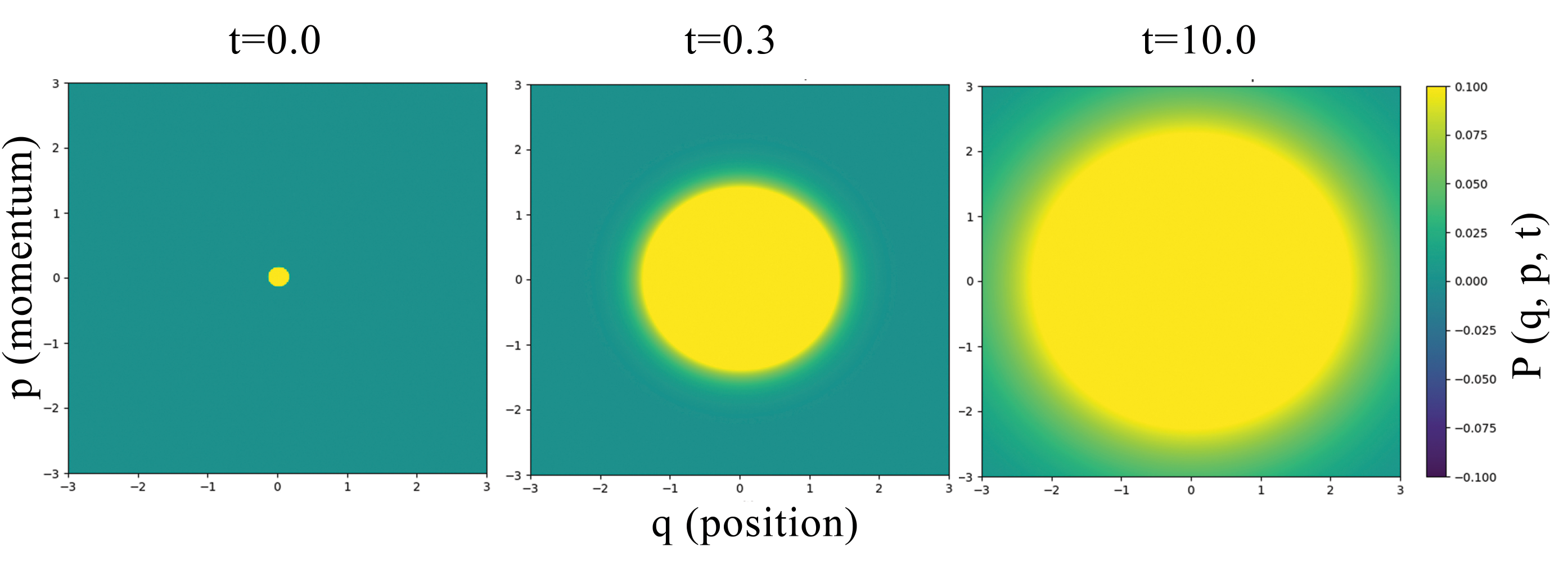}
    \caption{The evolution of the position and momentum dependent probability distributions, $(P(q,p,t))$, over time. Initially, the distribution is a Dirac delta function, transitioning to Gaussian probability distributions over time, illustrating the spreading from a localized to a dispersed state.}
    \label{fig:distribution}
\end{figure*}

\noindent In this section, we want to derive the $P$-representation of the quantum system that we have considered. For this system, we can write the master equation for the reduced density operator $\hat{\rho}_s$ 
from eq. \eqref{masteq} as
\begin{align}
	\frac{d\hat{\rho}_s}{dt}&=-\frac{\gamma_e+\gamma_c}{2}\left(\hat{a}^\dagger\hat{a}\hat{\rho}_s+\hat{\rho}_s\hat{a}^\dagger \hat{a}-2\hat{a}\hat{\rho}_s\hat{a}^\dagger\right)+ \left(\gamma_e\bar{n}_e+\gamma_c\bar{n}_c\right)\nonumber\\
	&\hspace{6mm}\times\left(\hat{a}\hat{\rho}_s\hat{a}^\dagger+\hat{a}^\dagger\hat{\rho}_s\hat{a}-\hat{\rho}_s\hat{a}\hat{a}^\dagger-\hat{a}^\dagger\hat{a}\hat{\rho}_s\right)~.\label{masteq2}
\end{align}
Now we represent the reduced density operator $\hat{\rho}_s $ in the coherent state representation as 
\begin{align}
	\hat{\rho}_s&=\int d^2\alpha ~ P(\alpha,\alpha^*,t)\ket{\alpha}\bra{\alpha}~.\label{fp1}
\end{align}
$P(\alpha,\alpha^*,t)$ in the above expression is the Glauber-Sudarshan quasi-probability distribution \cite{Glauber1963,sudarshan1963equivalence} and $\ket{\alpha}$ is the coherent state defined as \cite{LachsPaper}
\begin{align}
	\ket{\alpha}&=e^{-|\alpha|^2/2}\sum_{n=0}^{\infty}\frac{\alpha^n}{\sqrt{n!}}\ket{n}\label{fp2}
\end{align}
where $\ket{n}$ is the number state. \\
The action of creation and annihilation operator on the number state is given by \cite{sakurai2020modern,dirac1981principles}
\begin{subequations}\label{fp3}
	\begin{align}
		\hat{a}^\dagger\ket{n}&=\sqrt{n+1}\ket{n+1}~,\label{fp3a}\\
		\hat{a}\ket{n}&=\sqrt{n}\ket{n-1}~.\label{fp3b}
	\end{align}
\end{subequations}
Using eq.(s) (\eqref{fp2},\eqref{fp3}), we can write the following relations
\begin{subequations}\label{fp4}
	\begin{align}
		\hat{a}^\dagger\ket{\alpha}\bra{\alpha}&=\left(\frac{\partial}{\partial \alpha}+\alpha^*\right)\ket{\alpha}\bra{\alpha}~,\\
		\hat{a}\ket{\alpha}\bra{\alpha}&=\alpha\ket{\alpha}\bra{\alpha}~,\\
		\ket{\alpha}\bra{\alpha}\hat{a}^\dagger&=\alpha^*\ket{\alpha}\bra{\alpha}~,\\
		\ket{\alpha}\bra{\alpha}\hat{a}&=\left(\frac{\partial}{\partial \alpha^*}+\alpha\right)\ket{\alpha}\bra{\alpha}~.
	\end{align}
\end{subequations}
The above relations immediately lead to
\begin{align}
	&\hat{a}^\dagger\hat{a}\ket{\alpha}\bra{\alpha}+\ket{\alpha}\bra{\alpha}\hat{a}^\dagger \hat{a}-2\hat{a}\ket{\alpha}\bra{\alpha}\hat{a}^\dagger\nonumber\\
	&=\left(\alpha\frac{\partial}{\partial\alpha}+\alpha^*\frac{\partial}{\partial\alpha^*}\right) \ket{\alpha}\bra{\alpha}\label{fp5}
\end{align}
and
\begin{align}
	&\hat{a}\ket{\alpha}\bra{\alpha}\hat{a}^\dagger+\hat{a}^\dagger\ket{\alpha}\bra{\alpha}\hat{a}-\ket{\alpha}\bra{\alpha}\hat{a}\hat{a}^\dagger-\hat{a}^\dagger\hat{a}\ket{\alpha}\bra{\alpha}\nonumber\\
	&=-2\left(\alpha\frac{\partial}{\partial\alpha}+\alpha^*\frac{\partial}{\partial\alpha^*}+\frac{\partial^2}{\partial\alpha\partial\alpha^*}\right)\ket{\alpha}\bra{\alpha}\label{fp6}~.
\end{align}
Substituting eq. \eqref{fp1} in eq. \eqref{masteq2}, and using the results obtained in eq(s). (\eqref{fp5},\eqref{fp6}), and using the fact that the $P$ distribution vanishes in the boundary, we get
\begin{align}
	\frac{\partial P(\alpha,\alpha^*,t)}{\partial t}&=\frac{\gamma_e+\gamma_c}{2}\left(\alpha\frac{\partial}{\partial\alpha}+\alpha^*\frac{\partial}{\partial\alpha^*}\right) P(\alpha,\alpha^*,t)\nonumber\\
	&\hspace{6mm}+\left(\gamma_e+\gamma_c\right)\bar{n}_s\frac{\partial^2  P(\alpha,\alpha^*,t)}{\partial\alpha\partial\alpha^*}\label{fp7}
\end{align}
where $\bar{n}_s$ is given by eq. \eqref{steady_number}. Now we may separate out the real and imaginary parts of $\alpha$ as $\alpha=x+iy$, so that we can write $P(\alpha,\alpha^*,t)\equiv \tilde{P}(x,y,t)$. Then the Fokker-Planck equation can be written as
\begin{align}
\frac{\partial \tilde{P}}{\partial t}&=\frac{\gamma_e+\gamma_c}{2}\left(\frac{\partial}{\partial x}x+\frac{\partial}{\partial y}y\right) \tilde{P}\nonumber\\
	&\hspace{6mm}+\frac{\gamma_e\bar{n}_e+\gamma_c\bar{n}_c}{4}\left(\frac{\partial^2}{\partial x^2}+\frac{\partial^2}{\partial y^2}\right)\tilde{P}~.\label{fp8}
\end{align}
Separating the variables as
\begin{align}
\tilde{P}(x,y,t)=X(x,t)Y(y,t)~,\label{fp9}
\end{align}
the functions $X(x,t)$ and $Y(y,t)$ satisfy independent Fokker-Planck equations
\begin{subequations}\label{fp10}
\begin{align}
\frac{\partial X}{\partial t}&=\left(\frac{\gamma_e+\gamma_c}{2}\frac{\partial}{\partial x}x+\frac{\gamma_e\bar{n}_e+\gamma_c\bar{n}_c}{4}\frac{\partial^2}{\partial x^2}\right)X~,\\
\frac{\partial Y}{\partial t}&=\left(\frac{\gamma_e+\gamma_c}{2}\frac{\partial}{\partial y}y+\frac{\gamma_e\bar{n}_e+\gamma_c\bar{n}_c}{4}\frac{\partial^2}{\partial y^2}\right)Y~.
\end{align}
\end{subequations}
Using the standard procedure of solving Fokker-Planck equations \cite{fokker1914mittlere,planck1917satz,Carmichael,bhattacharjee2001statistical} with the initial conditions
\begin{subequations}\label{fp11}
\begin{align}
X(x,0|x_0,0)&=\delta(x-x_0)~,\\
Y(y,0|y_0,0)&=\delta(y-y_0)~,
\end{align}
\end{subequations}
we can write the solution for $X(x,t)$ as
\begin{align}
X(x,t|x_0,0)&=\frac{1}{\sqrt{\pi \bar{n}_s(1-e^{-\gamma t})}}\exp\left[-\frac{(x-x_0e^{-\gamma t/2})^2}{\bar{n}_s(1-e^{-\gamma t})} \right]\label{fp12}
\end{align}
where $\gamma=\gamma_e+\gamma_c$. Similarly the solution for $Y(y,t)$ can be written as
\begin{align}
Y(y,t|y_0,0)&=\frac{1}{\sqrt{\pi \bar{n}_s(1-e^{-\gamma t})}}\exp\left[-\frac{(y-y_0e^{-\gamma t/2})^2}{\bar{n}_s(1-e^{-\gamma t})} \right]\label{fp13}~.
\end{align}
Now we note that these $x$ and $y$ are the phase-space variables. For a mechanical oscillator these variables are given by the coordinate $q=x\sqrt{2\hbar/m\omega_s}$ and momentum $p=y\sqrt{2\hbar m\omega_s}$. Then the position dependent probability takes the form (from eq. \eqref{fp12})
\begin{align}
P(q,t)&=\frac{1}{\sqrt{\pi \bar{n}_s(1-e^{-\gamma t})}}\exp\left[-\frac{m\omega_s(q-q_0e^{-\gamma t/2})^2}{2\hbar\bar{n}_s(1-e^{-\gamma t})} \right]\label{fpe1}
\end{align}
and the momentum dependent probability takes the form (from eq. \eqref{fp13})
\begin{align}
P(p,t)&=\frac{1}{\sqrt{\pi \bar{n}_s(1-e^{-\gamma t})}}\exp\left[-\frac{(p-p_0e^{-\gamma t/2})^2}{2\hbar m\omega_s\bar{n}_s(1-e^{-\gamma t})} \right]\label{fpe2}
\end{align}
where $q_0$ and $p_0$ are the initial position and momentum of the system. It is reassuring to note that the distributions $P(q,t)$ and $P(p,t)$ will reduce to the Dirac delta function in the $t\to 0$ limit. From Fig. \ref{fig:distribution}, it is clear that initially the distribution is Dirac delta function, which gradually goes to a Gaussian probability distribution as time progresses. This illustrates the spreading of the probability distributions from a highly localized state to a more dispersed state.
\newline\noindent In the steady state, that is, when $t\to\infty$, the distributions in eq. (\eqref{fp12},\eqref{fp13}) can be written as
\begin{subequations}
\begin{align}
X(x)&=\frac{1}{\sqrt{\pi \bar{n}_s}}\exp\left[-\frac{x^2}{\bar{n}_s} \right]~,\\
Y(y)&=\frac{1}{\sqrt{\pi \bar{n}_s}}\exp\left[-\frac{y^2}{\bar{n}_s}\right]~.
\end{align}
\end{subequations}
Then the steady state Glauber-Sudarshan $P$-distribution can be written down as
\begin{align}
P(\alpha,\alpha^*)&=X(x)\cdot Y(y)\nonumber\\
&=\frac{1}{\pi\bar{n}_s}e^{-(x^2+y^2)/\bar{n}_s}\nonumber\\
&=\frac{1}{\pi\bar{n}_s}e^{-\left(m\omega_s q^2+\frac{p^2}{m\omega_s}\right)\big/2\hbar\bar{n}_s}\nonumber\\
&=\frac{1}{\pi\bar{n}_s}e^{-|\alpha|^2/\bar{n}_s}\label{fp14}
\end{align}
where $|\alpha|^2=\alpha^*\alpha=(x+iy)(x-iy)= x^2+y^2$. As a consistency check, from this probability distribution, we can again calculate the steady state number density as
\begin{align}
    tr\left(\hat{a}^\dagger\hat{a}\hat{\rho}_s\right)&=\frac{1}{\pi}\int d^2\beta~\bra{\beta}\hat{a}^\dagger\hat{a}\hat{\rho}_s\ket{\beta}\nonumber\\
    &=\frac{1}{\pi}\int d^2\beta ~ d^2\alpha~\bra{\beta}\hat{a}^\dagger\hat{a}P(\alpha,\alpha^*)\ket{\alpha}\braket{\alpha}{\beta}\nonumber\\
    &=\frac{1}{\pi^2\bar{n}_s} \int d^2\beta ~ d^2\alpha~\beta^*\alpha e^{-|\alpha|^2/\bar{n}_s} \left|\braket{\alpha}{\beta}\right|^2\label{fps1}
\end{align}
where we substituted $P(\alpha,\alpha^*)$ from eq. \eqref{fp14} in the third line of the above equality. From eq. \eqref{fp2}, and using the orthonormality condition for the number states, we have
\begin{align}
    \left|\braket{\alpha}{\beta}\right|^2&= e^{-\left|\alpha-\beta\right|^2}~.\label{fps2}
\end{align}
Substituting eq. \eqref{fps2} in eq. \eqref{fps1} gives
\begin{align}
    tr\left(\hat{a}^\dagger\hat{a}\hat{\rho}_s\right)&=\frac{1}{\pi^2\bar{n}_s} \int d^2\beta ~ d^2\alpha~\beta^*\alpha e^{-|\alpha|^2/\bar{n}_s} e^{-\left|\alpha-\beta\right|^2}\nonumber\\
    &=\bar{n}_s\label{fps3}~.
\end{align}
{From this probability distribution, we can now derive a quantity known as the entropic force \cite{verlinde2011origin,roos2014entropic,bhattacharya2023entropic}. An entropic force arises from the tendency of thermodynamic systems to maximize their entropy. It is a macroscopic force that emerges from the behavior of microscopic components within a system with numerous degrees of freedom, driven by the statistical inclination to increase entropy. The results clearly demonstrate the impact of the entropic force in the formation of a Bose-Einstein condensate for bosons and the manifestation of Pauli's exclusion principle for fermions.\\
Thus, from eq. \eqref{fpe1}, the entropic force reads}
\begin{align}
F(q,t)&=\frac{k_B T}{P}\frac{\partial P(q,t)}{\partial q}\nonumber\\
&=-\frac{m\omega_s k_B T}{\hbar\bar{n}_s}\frac{q-q_0e^{-\gamma t/2}}{1-e^{-\gamma t}}\label{fp16}
\end{align}
where $T$ is the temperature of the system. To interpret this temperature we need to look at the high temperature limit of the distributions. At high temperature, the distributions read
\begin{subequations}\label{fp17}
    \begin{align}
        \bar{n}_e&=\frac{1}{e^{\hbar\omega_s/k_B T_e}-1}\approx\frac{k_B T_e}{\hbar\omega_s}\\
        \bar{n}_s&=\frac{1}{e^{\hbar\omega_s/k_B T}-1}\approx\frac{k_B T}{\hbar\omega_s}\\
        \bar{n}_c&=\frac{1}{e^{\hbar\omega_s/k_B T_c}-1}\approx\frac{k_B T_c}{\hbar\omega_s}
    \end{align}
\end{subequations}
Putting these values in eq. \eqref{steady_number}, we can write
\begin{align}
    T=\frac{\gamma_eT_e+\gamma_cT_c}{\gamma_e+\gamma_c}\label{sys_temp}~.
\end{align}
This immediately shows $T_c<T<T_e$.\\
It is also interesting to observe that the entropic force in eq. \eqref{fp16} now has time dependence. From eq. \eqref{fp16}, we can write the entropic force in the steady state ($t\to\infty$ limit) as
\begin{align}
    F(q)&=-\frac{m\omega_s k_B T}{\hbar\bar{n}_s}q\label{fp18}~.
\end{align}
Substituting $\bar{n}_s$ (at high temperature) from eq. \eqref{fp17} in the above expression, we get
\begin{align}
    F(q)=-m\omega_s^2 q\label{fp19}~.
\end{align}
It is evident that the entropic force gives rise to the Hooke's law. This is in conformity with the fact that we are working with a bosonic harmonic oscillator system, and it is the entropic force that actually drives the interaction. It also confirms that the entropic force is exactly equal to the classical force as it was predicted in \cite{bhattacharya2023entropic}.
\section{Conclusion} \label{sec4}
\noindent The research undertaken here looks into complex dynamics about current and \textcolor{black}{quantum transport factor} in a bosonic system with a central unit coupled to two reservoirs at different temperatures. We formulate, in much detail, the Hamiltonian of the total system by carefully adding the bosonic interactions between the central system and the reservoirs. \textcolor{black}{The model considered in this work is physically relevant from the point of view of quantum transport.} We further present a thorough investigation of the time evolution of the central system's density matrix, by formally deriving a master equation allowing a more subtle current and \textcolor{black}{quantum transport factor} analysis.\\
Our investigation shows that the current, representing the flow of bosons in the system, is very strongly affected by the competition of the damping coefficients with the initial conditions. In the regime of the steady state, we derive an expression that characterizes the \textcolor{black}{quantum transport factor} of energy transfer in great detail. Remarkably enough, quantum effects due to temperature and the quantum correction factor strongly affect the total \textcolor{black}{transport factor} of the system.\\
We bring obvious improvements with our results. First, by using the concept of flux balance, we compute the current, which reproduces faithfully previous results. Then, we define a concept of \textcolor{black}{quantum transport factor}, from flux ratio, which incredibly leads to Carnot's {efficiency} as the leading term {when temperature goes to infinity and produces quantum corrections to the Carnot efficiency. Thus, the \textcolor{black}{quantum transport factor we obtain is greater than $\eta_c$ which is mathematically identical to the Carnot efficiency}}. This result is very important for quantum technological advancement.\\
Finally, we derive the {Fokker-Planck equation from} the master equation {using the Glauber-Sudarshan $P$-distribution for density matrix of the system. After solving the Fokker-Planck equation, we found a solution of the probability distribution. From the probability distribution in coordinate space, we find that the entropic force gives us the Hooke's law in the steady state, agreeing that the system of interest is a harmonic oscillator.}\\
\textcolor{black}{The investigation carried out in this work is also important from an applied point of view. In a recent work \cite{kansanen2024photon}, it has been observed that the statistics of transmitted photons in microwave cavities plays a central role in microwave quantum optics. A theory of photon counting statistics in Gaussian bosonic networks has been developed in this paper. It would therefore be natural to apply this method to the model considered in this work. We leave this for the future.} This paper also represents an important advance in the field of quantum thermodynamics and energy {\textcolor{black}{quantum transport factor}} processes in bosonic systems. The understanding gained has wide ramifications in fields like energy efficiency, where optimization of energy use is paramount. For sure, this will open new developments toward new quantum technologies that, again, will give unprecedented systems that are much more energy efficient.

\section*{Acknowledgment}
\noindent We would like to thank the referees for the useful comments made which has improved the quality of the paper substantially.

\bibliography{ref}

\end{document}